\documentclass{mem}
\usepackage{natbib}\usepackage{txfonts}\usepackage{balance}
\usepackage{graphicx}
\usepackage[a4paper]{hyperref}
\idline{75}{282}
\begin{document}
\def\teff{$T\rm_{eff }$}
\def\kms{$\mathrm {km s}^{-1}$}

\title{
The RR Lyrae distance scale from near-infrared photometry: current results
}

   \subtitle{}

\author{
M. \,Dall'Ora\inst{1} 
\and G. \, Bono\inst{2}
\and J. \, Storm\inst{3}
\and F. \, Caputo\inst{2}
\and G. \, Andreuzzi\inst{2,4}
\and G. \, Marconi\inst{5}
\and M. \, Monelli\inst{6}
\and V. \, Ripepi\inst{1}
\and P.~B. \, Stetson\inst{7}
\and V. \, Testa\inst{2}
          }

  \offprints{M. Dall'Ora}

\institute{
INAF - Osservatorio Astronomico di Capodimonte, via Moiariello 16,
I-80131 Naples, Italy; \email{dallora@na.astro.it, ripepi@na.astro.it}
\and
INAF - Osservatorio Astronomico di Roma, via Frascati 33, I-00040
Monte Porzio Catone (Rome), Italy;\email{bono@mporzio.astro.it,
testa@mporzio.astro.it}
\and
Astrophysikalisches Institut Potsdam, An der Sternwarte 16, D-14482
Potsdam, Germany; \email{jstorm@aip.de}
\and
Telescopio Nazionale Galileo, Istituto Nazionale di Astrofisica, P.O. Box
565, E-38700 Santa Cruz de La Palma, Spain; \email{andreuzzi@tng.iac.es}
\and
European Southern Observatory, 3107 Alonso de Cordova, Santiago,
Chile; \email{gmarconi@eso.org}
\and
IAC - Instituto de Astrofisica de Canarias, Calle Via Lactea E-38200 
La Laguna, Tenerife, Spain; \email{monelli@iac.es}
\and
Dominion Astrophysical Observatory, Herzberg Institute of 
Astrophysics, National Research Council, 5071 West Saanich Road, Victoria, 
British Columbia V9E 2E7, Canada; \email{Peter.Stetson@nrc-cnrc.gc.ca}
}
\authorrunning{Dall'Ora et al.}

\titlerunning{The RR Lyrae stars $PLK$ relation}

\abstract{We present new observational results on the RR Lyrae $K$-band
Period-Luminosity relation ($PLK$). Data on the Galactic globular clusters NGC 
3201 and NGC 4590 (M68), and on the Large Magellanic Cloud cluster Reticulum are
shown. We compare the observed slopes of the $PLK$ relations for these three
clusters with those predicted by pulsational and evolutionary models, finding a
fair  agreement. Trusting on this finding we decided to adopt these theoretical
calibrations to estimate the distance to the target clusters,
finding a good agreement with optical-based RR Lyrae distances, but with a
smaller formal scatter.

}
\maketitle{}

\section{Introduction}
RR Lyrae stars are well-known Population II variable stars, widely used as
distance indicators, since their mean $V$-magnitude is almost constant and they are
bright enough to be detected at moderately large distances. Usually their mean
$V$-magnitude is calibrated as a linear funtion of their abundance [Fe/H], 
but this calibration is still hampered by several theoretical and observational
uncertainties (see, e.g., \citealt{CacciariClementini} and
\citealt{Bonoreview}). However, in the $K$-band the RR Lyrae stars follow a 
quite tight Period-Luminosity relation ($PLK$, \citealt{l90}). This relation 
shows several advantages when compared to the optical calibration: the $PLK$ 
relation is only moderately affected by evolutionary effects, its intrinsic 
spread is of the order of 0.03 mag (\citealt{B01} (B01); \citealt{B03} (B03)), 
$K$-band magnitudes are only marginally affected by reddening, and the light 
curves have low amplitudes, allowing the estimate of the mean magnitude even 
with a few observations.
Moreover, empirical templates \citep{jones96} can be adopted to improve the
estimate of the mean magnitude when a single observation is available. However,
theoretical $PLK$ relations show a small dependence on the metallicity, with a 
slope of the order of $0.2$ mag/dex.
We present current results on the Galactic globular clusters
NGC 3201 and NGC 4590 (M68), and on the Large Magellanic Cloud (LMC) cluster
Reticulum. This work is part of a large observational project, which includes
data for 17 Galactic globular clusters, 5 LMC clusters, and field RR Lyrae
stars in the Baade Window, in the Sagittarius Stream, and in the main
body of the LMC.

\section{Observations and data reductions}
All the data presented in these proceedings were collected with SOFI@NTT and
pre-processed with IRAF routines following the prescriptions reported in the SOFI
user's manual. Photometry was carried out with
DAOPHOT/ALLFRAME packages (\citealt{stetson87}; \citealt{stetson94}).  
The photometric zero-points were computed from
the standards catalogue by \citet{persson}, which is defined in the LCO system. 
For NGC 3201 and M68, a cross-check with local standards from the 2MASS 
catalogue showed a difference in magnitude of $0.01$ mag in the
$K_s$-band, in excellent agreement with the shift predicted by \citet{carpenter}
between these two NIR systems. To properly compare empirical results and theoretical
predictions, computed in the \citet{besselbrett} system, we finally transformed
observations and theory to the 2MASS system following the equations by
\citet{carpenter}.

\begin{figure}[]
\resizebox{\hsize}{!}{\includegraphics[clip=true]{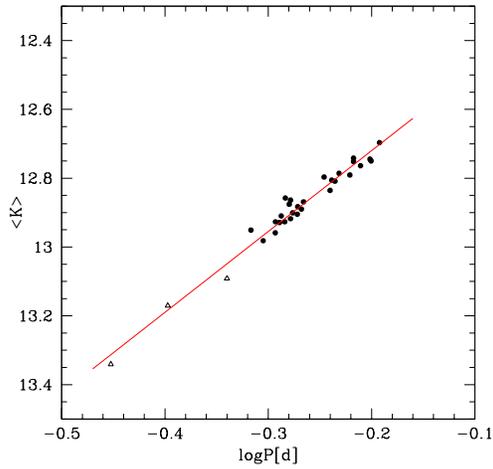}}
\caption{
\footnotesize
Observed $PLK$ relation for NGC 3201. Filled circles are fundamental pulsators,
while filled triangles depict first overtone fundamentalized RR Lyrae stars.
The straight line marks the empirical fit to the data.
}
\label{n3201plk}
\end{figure}

NGC 3201 is a metal-intermediate cluster ([Fe/H]$=-1.53$), with a large number 
of RR Lyrae stars (77) and characterized by a strong
differential reddening (\citealt{laydensarajedini}; \citealt{piersimoni3201}), 
that makes it difficult the estimate of the distance from $V$-band
photometry. Owing to its large extent, this cluster was observed with two 
different pointings, covering only the southern part. 
Therefore, only 30 RR Lyrae stars are present in our catalogue, for which we
collected a dozen phase points. Observed light
curves were fitted with cubic splines, and intensity-averaged mean magnitudes 
were computed. Figure \ref{n3201plk} shows the observed $PLK$ relation for
this cluster. The straight line marks the empirical fit to the complete set of
RRab (27) and fundamentalized ($\log P_F = \log P + 0.127$) RRc stars (3), 
with a slope of $-2.35 \pm 0.08$. This slope
is slightly steeper than the slope predicted by the pulsational models (-2.101,
B03; -2.071, B01), but in close agreement with the slopes based on evolutionary 
models ($\sim -2.34$, \citealt{marcio}; \citealt{santino}).
When only RRab stars are used, the observed slope is $-2.13 \pm 0.12$.
Using the fully theoretical B01 calibration, and an average reddening of
$E(B-V)=0.30$ mag \citep{piersimoni3201}, the true distance modulus is $13.38
\pm 0.03$ mag, in good agreement with the optical estimates
\citep{laydensarajedini}. The semi-empirical calibration by B03 gives a longer
distance of $13.47 \pm 0.03$ mag. 

\begin{figure}[]
\resizebox{\hsize}{!}{\includegraphics[clip=true]{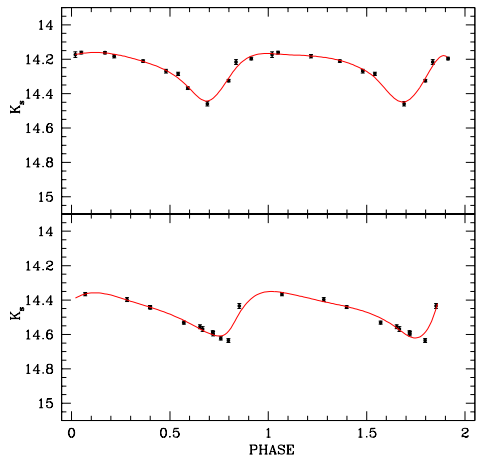}}
\caption{
\footnotesize
Light curves for the M68 RR Lyrae stars V10 (bottom panel) and V35. Observations
were fitted with a cubic spline (solid line).
}
\label{step}
\end{figure}

M68 is a metal-poor cluster ([Fe/H]$=-2.1$),
characterized by a sizable sample of
RR Lyrae stars (42) and by a well-known reddening ($E(B-V)=0.07 \pm 0.01$,
\citealt{walker94}). We collected data for two different pointings, recovering 34 RR Lyrae stars. 
Observed light curves
(12 phase points) were fitted with cubic splines (see Figure \ref{step} for an
example). 
M 68 contains a large number of suspected Blazhko stars and double pulsators, and for 
these variables the light curves show a large spread. Mean magnitudes for these variables 
could be improved with the use of the 
templates. 

\begin{figure}[]
\resizebox{\hsize}{!}{\includegraphics[clip=true]{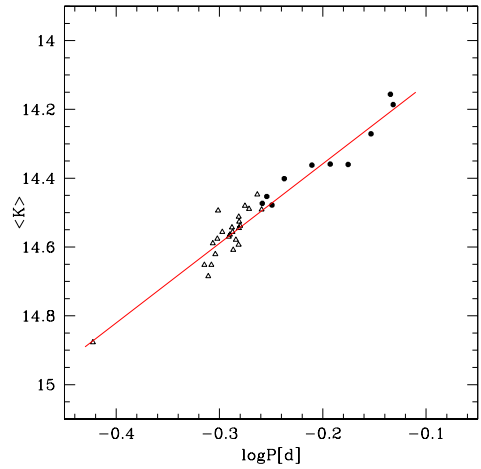}}
\caption{
\footnotesize
Observed $PLK$ relation for NGC 4590. Filled circles are fundamental pulsators,
while open triangles depict first overtone "fundamentalized" RR Lyrae stars.
The straight line marks the empirical fit to the data.
}
\label{m68plk}
\end{figure}

Figure \ref{m68plk} shows the empirical $PLK$ relation for M 68. Filled circles
represent RRab stars, while empty triangles show the position of fundamentalized
RRc stars. The straight line depicts the empirical fit to the data. The observed
slope, when the whole sample (RRab + RRc stars) is considered, is $-2.36 \pm
0.11$, in good agreement with the prediction of the evolutionary models of Cassisi
et al. (2004). However, if only the fundamental pulsators are used, the observed
slope is $-2.19 \pm 0.23$. Adopting the theoretical calibration of B01, the
true distance is $15.11 \pm 0.04$, in excellent agreement with
the most recent estimates available in the literature \citep{dicrisci}. As in
the case of NGC 3201, the B03 calibration yields a larger distance ($\mu_0=15.24$
mag).

Data presented for Reticulum have already been discussed in \citet{dallora_reti}.
Here we focus our attention on the estimated distance: adopting the
B01 calibration, we estimate a true distance of $18.46 \pm 0.03$ mag, 
slightly smaller than the distance estimated with the B03 calibration, 
$18.52 \pm 0.03$ mag, adopted in \citet{dallora_reti}. 

The observed slopes and the estimated distances are summarized in Tables 1 and 2, 
respectively.

\begin{figure}[]
\resizebox{\hsize}{!}{\includegraphics[clip=true]{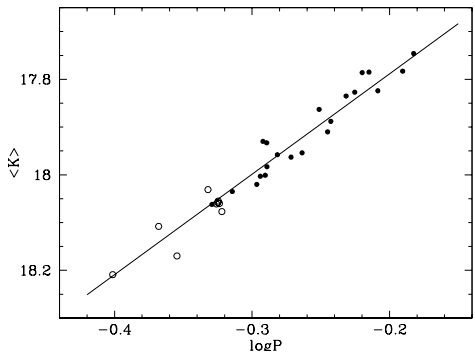}}
\caption{
\footnotesize
Observed $PLK$ relation for the LMC cluster Reticulum. Filled and open circles
show fundamental and fundamentalized pulsators, respectively. The straight
line marks the theoretical prediction from B03 for the derived distance modulus.
Plot from \citet{dallora_reti}.
}
\label{reti}
\end{figure}

\begin{table*}[]
\caption{Observed and predicted slopes. See text for details.}
\label{slopes}
\begin{center}
\begin{tabular}{lcccccc}
\hline
Cluster & RRab + RRc & RRab & B01 & B03 & Catelan et al. & Cassisi et al. \\
\hline
NGC 3201  &$ -2.35 \pm 0.08 $ & $ -2.13 \pm 0.12 $ & $ $ &$ $ & $-2.345$ & $-2.34 $\\
M 68      &$ -2.36 \pm 0.11 $ & $ -2.19 \pm 0.23 $ & $ -2.071 $ &$-2.101$ & $...$ & $-2.30 $\\
Reticulum &$ -2.16 \pm 0.09 $ & $ -2.19 \pm 0.11 $ & $ $ & $ $ & $-2.358$ & $-2.34  $ \\
\hline
\end{tabular}
\end{center}
\end{table*}

\begin{table*}[]
\caption{Distances: we list optical-based RR Lyrae distances
(NGC 3201:\citealt{laydensarajedini}; M 68: \citealt{dicrisci}; Reticulum: \citealt{walker92}),
and the estimated true distances with the current calibrations of the $PLK$.}
\label{slopes}
\begin{center}
\begin{tabular}{lccccccc}
\hline
Cluster & [Fe/H] & Optical & B01 & B03 & Catelan et al. & Cassisi et al. \\
\hline
NGC 3201  &$ -1.53 $ & $13.40 \pm 0.05$ & $13.38 \pm 0.03 $ & $13.47 \pm 0.03$ & $13.27 \pm 0.03$ & $13.38 \pm 0.02 $\\
M 68      &$ -2.10 $ & $15.08 \pm 0.07$ & $15.11 \pm 0.04 $ & $15.24 \pm 0.06$ & $...$ & $15.15 \pm 0.04 $\\
Reticulum &$ -1.71 $ & $18.35 \pm 0.10$ & $18.46 \pm 0.03 $ & $18.52 \pm 0.03$ & $18.31 \pm 0.03$ & $18.47 \pm 0.03$\\
\hline
\end{tabular}
\end{center}
\end{table*}

\section{Discussion and conclusions}
The observed slopes for the three targets are listed in Table 1. Current slopes
have been estimated using the entire sample of variables and only for fundamental
variables. The slopes predicted by pulsational and evolutionary models are also
listed.
In Table 2 we compare the estimated distances using the four most recent
calibrations of the $PLK$ available in the literature, and for comparison the
optical-based distances. Values listed in this table indicate that B03 calibration gives
longer distances than B01. This difference can be accounted for with the metallicity
coefficient ($+0.167$ and $+0.231$ in the B01 and B03 calibrations,
respectively). Finally, we note that $PLK$ distances are characterized by a 
smaller formal error than optical
distances. It is worth noting that the $PLK$ relations estimated by \citet{santino} 
and by \citet{marcio} require knowledge of the HB
morphological type. This parameter is completely unknown for
field RR Lyrae stars. On the other hand, pulsationally based $PLK$ relations
only need knowledge of the metallicity and, once calibrated, can be adopted
for individual RR Lyrae stars. 
To this aim, we have already collected accurate near-infrared data for a
large sample of globular clusters that cover a wide range of metal contents
and HB morphologies, and firmer conclusions will be drawn when the 
photometry of this dataset is completed.

\bibliographystyle{aa}

\end{document}